\documentclass[conference]{IEEEtran}
\usepackage{stmaryrd}
\usepackage{amsfonts}
\usepackage{bbm}
\usepackage{graphicx,times,psfig,amsmath, url} % Add all your packages here
\usepackage{multirow}

\IEEEoverridecommandlockouts    % to create the author's affliation portion
\begin{document}

% paper title: Must keep \ \\ \LARGE\bf in it to leave enough margin.
\title{\ \\ \LARGE\bf Parallel multi-objective algorithms for the molecular docking problem \thanks{Jean-Charles Boisson, Laetitia Jourdanand El-Ghazali Talbi are members of the INRIA project team DOLPHIN, 40, avenue Halley 59650 Villeneuve d'Ascq, France. Dragos Horvath is in a CNRS laboratory, B{\^a}t. C9 Cit{\'e} Scientifique 59655 Villeneuve d'Ascq, France. Email: \{Jean-Charles.Boisson, Laetitia.Jourdan, El-Ghazali.Talbi\}@lifl.fr, Dragos.Horvath@univ-lille1.fr}\thanks{This work was supported by the ANR DOCK project and the PPF BioInfo of Lille1.}}

\author{Jean-Charles Boisson, Laetitia Jourdan, El-Ghazali Talbi and Dragos Horvath}
% avoiding spaces at the end of the author lines is not a problem with
% conference papers because we don't use \thanks or \IEEEmembership
% use only for invited papers
%\specialpapernotice{(Invited Paper)}

% make the title area
\maketitle

\begin{abstract}
  Molecular docking is an essential tool for drug design. It helps the scientist to rapidly know if two molecules, respectively called 
ligand and receptor, can be combined together to obtain a stable complex. We propose a new multi-objective model combining an 
energy term and a surface term to gain such complexes. The aim of our model is to provide complexes with a low energy and low surface. 
This model has been validated with two multi-objective genetic algorithms on instances from the literature dedicated to the 
docking benchmarking. 
\end{abstract}

% no key words

\section{Introduction}

\PARstart{F}{or} drug design, it is essential to find which molecules can interact with other bigger molecules. In this context, the docking problem
consists in finding how a small molecule, the ligand, can be put in contact in a particular location, the binding site, of another
bigger molecule. Experimental docking studies cost time and resources. There generally exist more than one hundred thousand
ligands and the binding site of a receptor is not necessary known and/or unique. In this situation, automatic docking methods
to screen large ligand databases allow to speed up drug design. The ligand databases are parsed in order to find ligands which can be 
docked with the molecule of interest in order to enable, disable or modify its function. Then the selected ligands can be docked 
experimentally to validate the result of the automatic docking. These approaches to speed-up drug design are also called 
``virtual screening'' methods. 

Since the 90's, metaheuristics have been used to solve the molecular docking problem. Originally, single solution metaheuristics, such as 
Metropolis Monte-Carlo algorithm or Simulated Annealing, were used to solve this problem. For example, the first version of the 
well known AutoDock software package has its main algorithm based on a Simulated Annealing~\cite{BGO1990}. Later, population based 
metaheuristics like Genetic Algorithms (GAs) have been used~\cite{BJW1995,BY2003}. The current main algorithm included in AutoDock is 
based on a Lamarckian Genetic Algorithm (LGA). It corresponds to the hybridization of a GA and a local search method~\cite{BMG1998}. 
Recently, new docking methods have been also proposed using Particle Swarm Optimization (PSO)~\cite{BJMM2007} or Ant Colony based 
metaheuristics~\cite{BKS2007}. All these methods try to find the best binding mode using complete molecules. Other methods propose 
incremental algorithms to find the binding mode. In DOCK~\cite{BEK1997} and FlexX~\cite{BRK1996}, the complete ligand is constructed 
step by step in the binding site. More information about standard docking softwares can be found in~\cite{BBT2003}. 

We propose a new multi-objective model for the flexible docking problem combining an energy term with a surface term. It is a 
flexible docking model because the conformation of the ligand and the site can be modified during the process. The aim of the surface
 term is to guide the penetration of the ligand into the site. The energy term is used to gain a complex of low energy.  

This paper is organized in four main parts. First, our bi-objective model is detailed and each objective is presented in three 
steps: definition, motivation and validation. In the second part, the algorithm design is described. As we use a platform to ease the design of our 
algorithm, only parts dedicated to the docking problem are explained. The third part presents our first results that validate our model. 
Finally, conclusions and perspectives about this work are provided.

\section{New model for the molecular docking problem}

\subsection{Existing multi-objective models}

Most of the docking methods use a mono-objective modeling. In these models, the objective is generally the binding free energy. This 
objective is defined as an aggregation of energy interaction terms. However, other type of information can be also included. First 
multi-objective models were based on subsets of the original binding free energy from the mono-objective models. The multi-objective 
model that is the most used for solving the docking problem (but also the protein structure prediction problem) is the bi-objective model 
that divides the energy into bonded and non-bonded energy. This model is based on the notion of attractive and repulsive energies that 
maintain the molecule into a stable conformation. Other models include objectives based on information about molecule 
geometry~\cite{BOT2006}. But this type of objective is more often used in preliminary studies for decreasing the search space of docking
 methods~\cite{BYM2006}. 

\subsection{Our bi-objective model}

In our model, we combine an energy term and a surface term. The first one describes the stability of the ligand/site complex (LSC) and
the second allows to qualify the how the ligand is entered into the binding site.

\subsubsection{First objective}

This criterion is a compound of two main terms: the bonded and the non-bonded atom energy. The first describes all the 
interactions that occur when two atoms are linked together. This term is described in equation~\ref{EnergyPartOne}. 

\begin{equation}
  \label{EnergyPartOne}
  \begin{split}
    E_{bonded\_atoms}~= & \sum_{bonds}{K_b(b-b_0)^2}~+\\
    & \sum_{angles}{K_{\theta}(\theta-\theta_0)^2}~+\\
    & \sum_{torsions}{K_{\phi}(1-cos~n(\phi-\phi_0))}
  \end{split}
\end{equation} 

$K_b$, $K_{\theta}$ and $K_{\phi}$ are the strength constants linked to the length, the angle and the phase contributions respectively.
In the same manner, $b_0$, $\theta_0$ and $\phi_0$ are empirical optimal value for the length, the angle and the phase difference 
between two given atoms. $b$, $\theta$ and $\phi$ are the current values of the length, the angle and the phase difference. For 
the torsion term, $n$ is the periodicity linked to the type of the central bond of the torsion (double or triple).

The second term of our first objective function corresponds to the interactions between the atoms and their environment (other atoms, 
solvent, etc). This term is detailed in equation~\ref{EnergyPartTwo}.

\begin{equation}
  \label{EnergyPartTwo}
  \begin{split}
    E_{non\_bonded\_atoms}~= & \sum_{Van~der~Waals}{\frac{K^a_{ij}}{d^{12}_{ij}}-\frac{K^b_{ij}}{d^6_{ij}}}~+\\
    & \sum_{Coulomb}{\frac{q_iq_j}{4 \pi \epsilon d_{ij}}}~+\\
    & \sum_{desolvation}{\frac{Kq^2_iV_j+q^2_jV_i}{d^4_{ij}}}
  \end{split}
\end{equation} 

In this equation, $q_i$ is the charge of the atom $i$; $d_{ij}$ is the distance between atoms $i$ and $j$; $V_i$ is a volumetric 
measure for the atom $i$; $K$ and $K_{ij}^x$ are strength constants linked to the contribution of the considered atoms. The Van der
 Waals contribution term allows to describe the combination of attractive and repulsive force between two atoms according to the distance
between their centers. The Coulomb contribution term describes how the electronegativity differences inside a molecule between atoms 
of different size and mass have an impact on the corresponding energy. These differences produce charges that can be attractive or 
repulsive. The desolvation term models the solvent action around a molecule. 

The force field used for computing all these terms is the Consistent Valence Force Field (CVFF). All the parameters of this
force field have been tuned experimentally on a diverse set of molecules.

These bonded and non-bonded energy terms have been already used in a bi-objective model for the resolution of the Protein 
Structure Prediction problem (PSP)~\cite{BTMT2007}. %In~\cite{BJM2005} the authors describe a bi-objective PSO using the inter and the
%intra energy for the docking problem.

In our case, the first criterion is a stability indicator. To estimate the stability of a ligand/site complex, we need its 
complete molecular energy. As a result, the bonded and the non-bonded energy terms are combined. Finally, our first objective function 
is a compound of six terms summarized in equation~\ref{Energy}:

\begin{equation}
  \label{Energy}
  \begin{split}
  E~= &~E_{bonded\_atoms}~+~E_{non\_bonded\_atoms}\\ 
  = & \sum_{bonds}~+~\sum_{angles}~+~\sum_{torsions}~+\\
  & \sum_{Van~der~Waals}~+~\sum_{Coulomb}~+~\sum_{desolvation}
  \end{split}
\end{equation} 

Our first criterion defines the molecular energy of a LSC. The lower the energy is, the more stable the complex is. Nevertheless, 
a LSC with a low energy does not necessarily correspond to a good quality docking. Two LSCs with an equivalent energy can 
correspond to two completely different complexes. When considered alone, energy cannot give enough information to differentiate similar 
conformations. A same level of energy can correspond to a very diversified family of conformations. A family of narrow conformations, 
can have very different levels of energy. Our second objective function may help choosing the best LSC for 
our problem.

\subsubsection{Second objective}

For molecules, there are three types of surfaces:

\begin{itemize}
\item{the Van Der Waals Surface (VDWS) that is the simplest surface to represent.}
\item{the Solvent Accessible Surface (SAS) that is the first to use the notion of solvent.}
\item{the Connolly Surface (CS) that is considered as the real surface of a molecule.}
\end{itemize}

\begin{figure}[h]
  \centering{\includegraphics[scale=0.5]{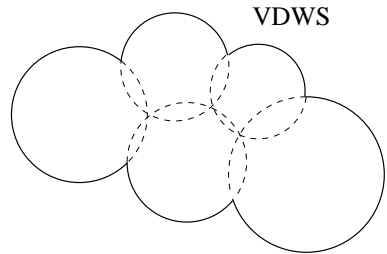}}
  \caption{Van der Waals surface (VDWS) of a molecule compound of five atoms.}
  \label{fig:SASA_VDW}
\end{figure}

An atom can be represented as a sphere due to its Van der Waals radius. The VDWS corresponds to the sum of the spherical 
surface parts that are not in collision with other spheres. Figure~\ref{fig:SASA_VDW} shows the Van der Waals surface of a molecule of
five atoms. 

\begin{figure}[h]
  \centering{\includegraphics[scale=0.5]{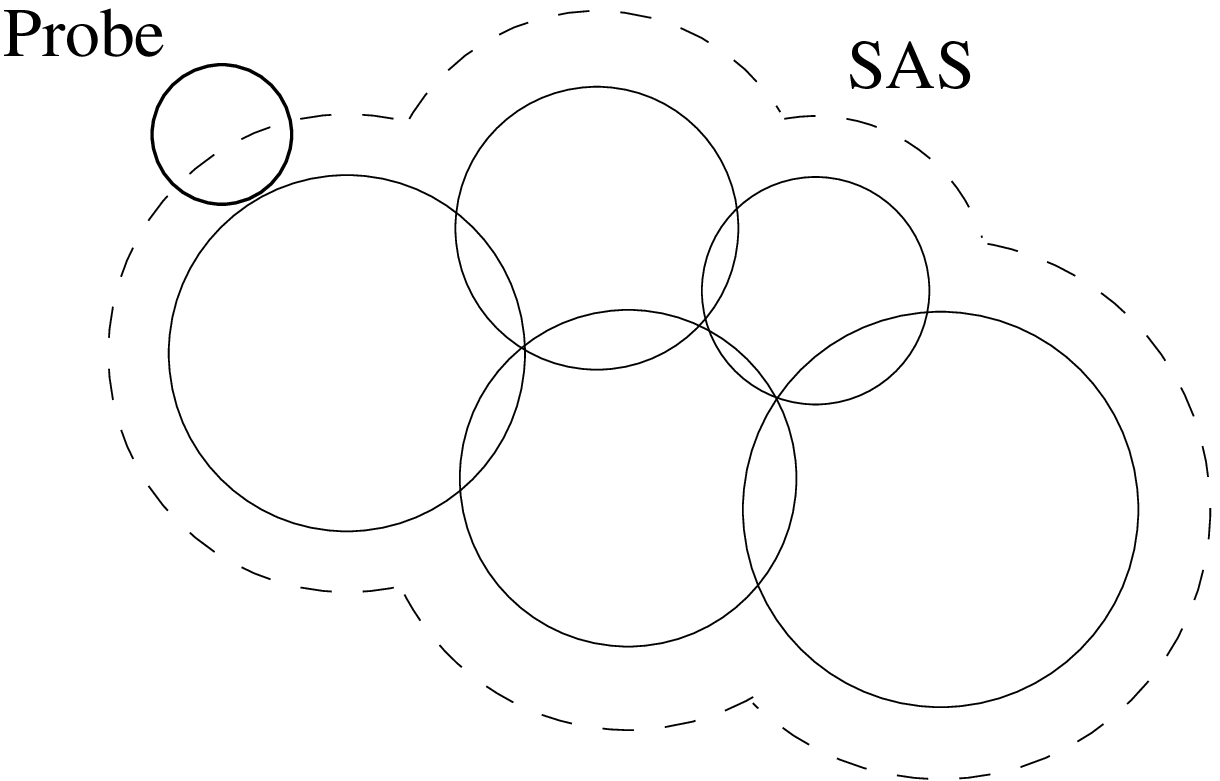}}
  \caption{Solvent accessible surface (SAS) of a molecule compound of five atoms. The probe symbolizes a solvent molecule. In our case, it is a molecule of water.}
  \label{fig:SASA_SAS}
\end{figure}

The SAS, and later the CS, were defined by Lee and Richards in~\cite{BLR1971} and in~\cite{BR1977} respectively. For the VDWS, 
the molecule is considered to be in the vacuum but it is a simplified model. The SAS and the CS are more realistic 
surfaces because they consider that the molecule is in a solvent. This solvent presence is symbolized by a probe. 
The SAS is drawn according to the center of this probe that rolls on the atom spheres. Generally, the probe has a radius 
of 1.4~\AA~(1 angstrom(\AA)= 0.1 nanometer) in order to be able to contain a water molecule that is one of the standard 
solvents. Figure~\ref{fig:SASA_SAS} describes the SAS of the same molecule of five atoms.

\begin{figure}[h]
  \centering{\includegraphics[scale=0.5]{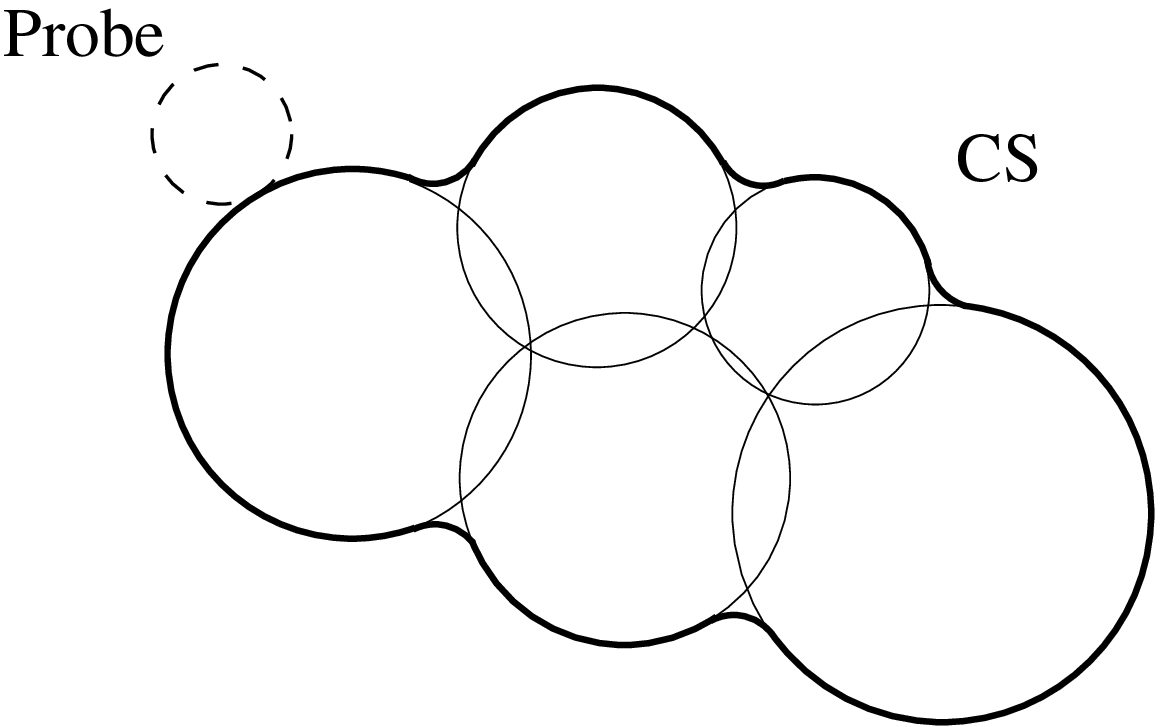}}
  \caption{Connoly Surface (CS) of a molecule compounded of five atoms. The probe symbolizes a solvent molecule. In our case, it 
is a molecule of water.}
  \label{fig:SASA_CS}
\end{figure}

For the CS, the surface is drawn according to all the points of the probe surface that touch the atom spherical surfaces. A special 
case occurs when a probe touches two spheres at the same time. In this case, the drawn surface corresponds to all the points 
of the probe surface which are oriented toward the molecule. An example of CS is shown in the figure~\ref{fig:SASA_CS} always with
the same molecule of five atoms.

Several methods that compute these surfaces can be found in~\cite{BFB1998, BFAR2000, BRPK2006, BVH1997}.

For our multi-objective model, we use an algorithm that computes an approximation of the SAS for a LSC.
The SAS is a good compromise between quality and computational complexity. Due to the notion of solvent, it is a realistic surface
and its calculus is not too expensive compared to the CS computation. The original SAS algorithm was first presented 
in~\cite{BGM1993}, but was also recently used in~\cite{BLFB2007}. It is based on look-up tables and Boolean Logic. It approximates the 
method of Shrake and Rupley~\cite{BSR1973}. 

\begin{figure}[h]
  \centering{\includegraphics[scale=0.5]{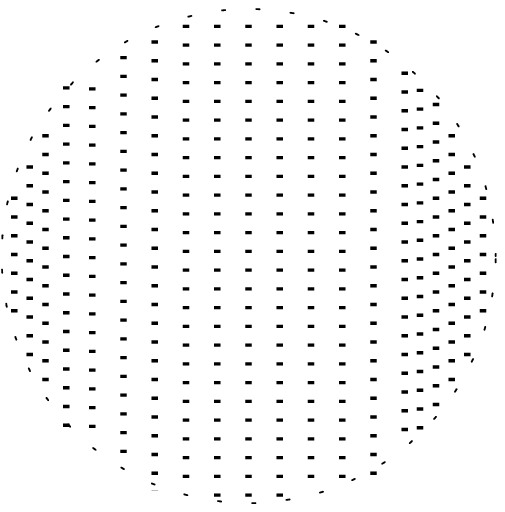}}
  \caption{Representation of the spherical surface of an atom with points.}
  \label{fig:ATOM_SURFACE}
\end{figure}

According to this method, each atom spherical surface is represented as a set of points (figure~\ref{fig:ATOM_SURFACE}). 
Each point is encoded as a bit to indicate if it is in interaction with the solvent (1) or not (0). Thus surface points are represented 
as bit string. Due to the atom encoding, computing the area of one atom only consists of ``AND'' Boolean operations. Look-up tables are 
used to speed-up the calculus by saving Boolean masks used to approximate intersection between atom points. The SAS of a LSC allows to 
evaluate the penetration of the ligand into the site. In a real docking process, the ligand may try to dive into the binding site or try 
to modify its conformation to better suit the binding surface. In both cases, the corresponding SAS will decrease. Therefore, this 
criterion is essential for simulating realistic flexible docking processes. 

\section{Method}

\subsection{Multi-objective optimization problems}

In a variety of applications, a problem arises that several objective functions have to be optimized concurrently. 
One important feature of these problems is that the different objectives typically contradict each other and therefore 
certainly not have identical optima. Thus, the question arises how to approximate one or several particular ``optimal compromises'' or 
how to compute all optimal compromises of this multi-objective optimization problem (MOP).

A MOP can be defined as follow:
$$\min_{x \in S}\{F(x)\},~S=\{x\in\mathbbm{R}^n:h(x)=0, g(x)\leq 0\}, $$
where $F$ is defined as the vector of the objectives:
$$F:\mathbbm{R}^n\rightarrow\mathbbm{R}^k,~F(x)=(f_1(x), ..., f_k(x)),$$
with $f_1, ...,f_k:\mathbbm{R}^n\rightarrow\mathbbm{R}, h:\mathbbm{R}^n\rightarrow\mathbbm{R}^m, m \leq n,$ and 
$g:\mathbbm{R}^n\rightarrow\mathbbm{R}^q$. A vector $v\in\mathbbm{R}^k$ is said to be \emph{dominated} by a vector $w\in\mathbbm{R}^k$
if for all $i\in{1, ...,k}$ $w_i \leq v_i$ and $v \neq w$. A vector $v$ is \emph{nondominated} with respect to a set $P$, if none
of the vectors $p\in P$ dominate $v$. A point $x\in S$ is called optimal or \emph{Pareto optimal}, if $F(x)$ is not dominated by any 
vectors $F(y), y\in S$. 

\subsection{ParadisEO platform}

In order to ease the implementation of our algorithm, we have used the ParadisEO platform~\cite{CCMT2004}. ParadisEO is a 
complete platform to design powerful optimization methods. It consists in four components:
\begin{enumerate}
\item{ParadisEO-\textbf{EO} (\textbf{E}volving \textbf{O}bject) dedicated to population-based metaheuristics.}
\item{ParadisEO-\textbf{MO} (\textbf{M}oving \textbf{O}bject) dedicated to single solution-based metaheuristics.} 
\item{ParadisEO-\textbf{MOEO} (\textbf{M}ulti \textbf{O}bjective \textbf{EO}) dedicated to multi-objective meta-heuristics.}
\item{ParadisEO-\textbf{PEO} (\textbf{P}arallel \textbf{EO}) dedicated to parallel metaheuristics.}
\end{enumerate}
 
ParadisEO-MOEO~\cite{CLMJ2007} and ParadisEO-PEO have been more particularly used in our case. More information
about ParadisEO is available on the official website:

$$ (http://paradiseo.gforge.inria.fr).$$

This platform allows the user to only design the parts specific to his problem in order to design effective algorithms. In our case, 
only solution encoding, solution evaluation and genetic operators have implemented.

\subsection{Parallel genetic algorithms}

Thanks to the ParadisEO platform, two parallel genetic algorithms have been designed: one based on the well known NSGA-II 
(\emph{Non-dominated Sorting Genetic Algorithm II}) and the other on the IBEA (\emph{Indicator-Based Evolutionary Algorithm}).
The first one is a standard multi-objective algorithm used to test our model. The second one is an algorithm that has been proved
better than NSGA-II on several problems. Therefore, we have test it on the docking problem. 

\subsubsection{Genetic Algorithms}

A Genetic Algorithm (GA) works by repeatedly modifying a population of artificial structures through the application of genetic 
operators (crossover and mutation)~\cite{hollan:ga}. The goal is to find the best possible solution or, at least good, solutions for 
the problem.

\subsubsection{NSGA-II and IBEA}

In NSGA-II~\cite{CDPA2002}, the solutions contained in the population are ranked into several classes at each 
generation. Individuals from the first front all belong to the first efficient set. Individuals from the second front all belong to 
the second best efficient set, etc. Two values are then computed for every solutions of the population. The first one corresponds 
to the rank the corresponding solution belongs to, and represents the quality of the solution in terms of convergence. The second one, 
the crowding distance, consists of estimating the density of solutions surrounding a particular point of the objective space, and 
represents the quality of the solution in terms of diversity. A solution is said to be better than another if it has the best rank, 
or in the case of a tie, if it has the best crowding distance. The selection strategy is a deterministic tournament between two random 
solutions. At the replacement step, only the best individuals survive, with respect to the population size. Likewise, an external 
population is added to the steady-state NSGA-II in order to store every potentially efficient solution found during the search.

For IBEA~\cite{CZK2004}, the fitness assignment scheme is based on a pairwise comparison of solutions contained in a population by 
using a binary quality indicator. No diversity preservation technique is required, according to the indicator being used. The selection 
scheme for reproduction is a binary tournament between randomly chosen individuals. The replacement strategy is an environmental 
one that consists of deleting, one-by-one, the worst individuals, and in updating the fitness values of the remaining solutions each 
time there is a deletion; this is continued until the required population size is reached. Moreover, an archive stores solutions 
mapping to potentially non-dominated points, in order to prevent their loss during the stochastic search process. 

\subsubsection{Coding}

\begin{figure}[h]
  \centering{\includegraphics[scale=0.5]{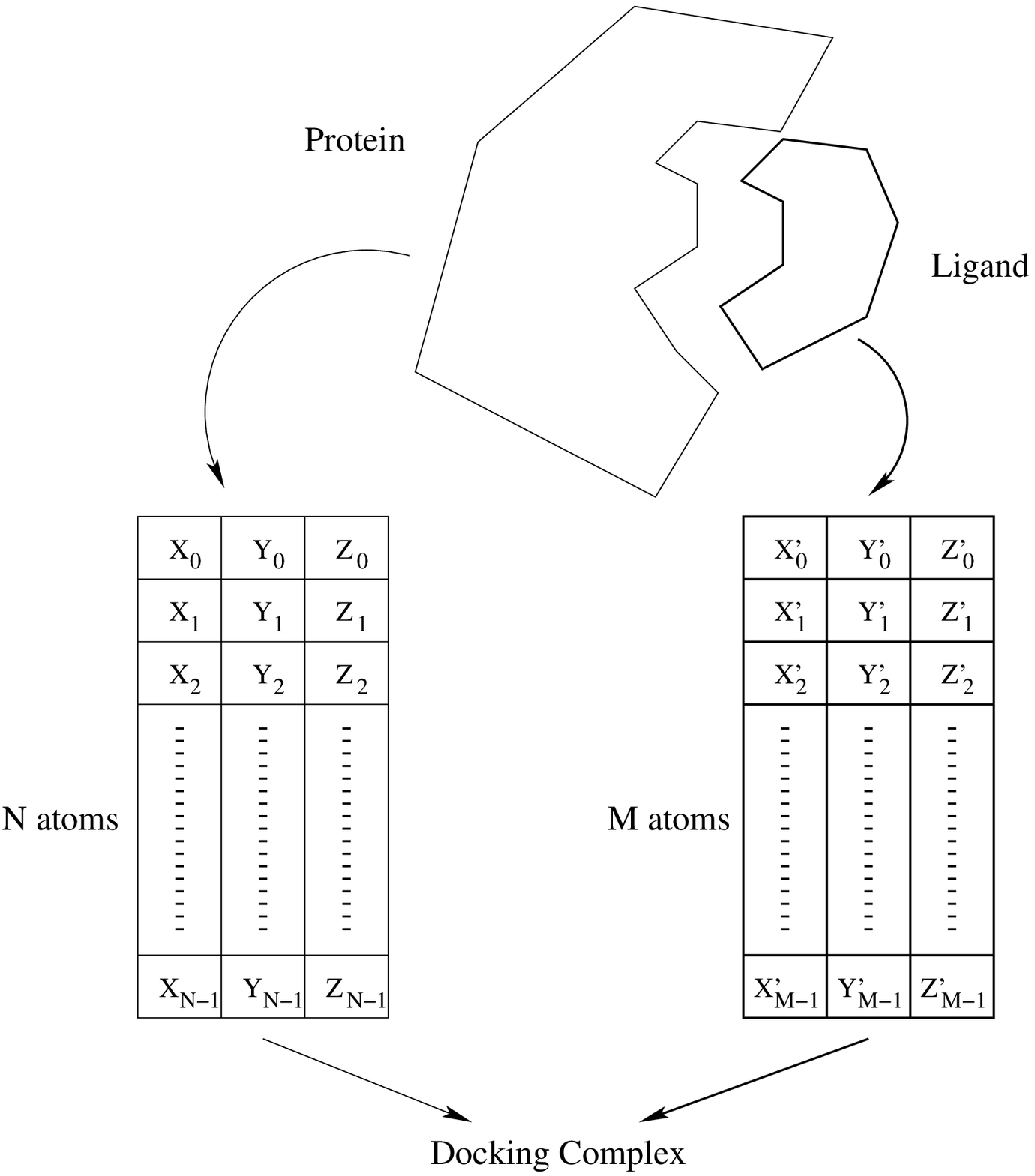}}
  \caption{Representation a solution in our genetic algorithm. N and M are the number of atoms compounding the binding site and the ligand respectively.}
  \label{fig:individualCoding}
\end{figure}

In our algorithm, the solutions are represented according to two vectors of float corresponding to the atomic coordinates. Each atom
has three coordinates (x, y and z). Figure~\ref{fig:individualCoding} describes this coding. In our case a solution is 
called a ``Docking Complex''.

Between two individuals, only the coordinates of the atoms change. The molecule topology is already loaded and can be used directly. 
The full ligand/site complex is only build during the evaluation step of an individual. 

\subsubsection{Operators}

\begin{figure}[h]
  \centering{\includegraphics[scale=0.5]{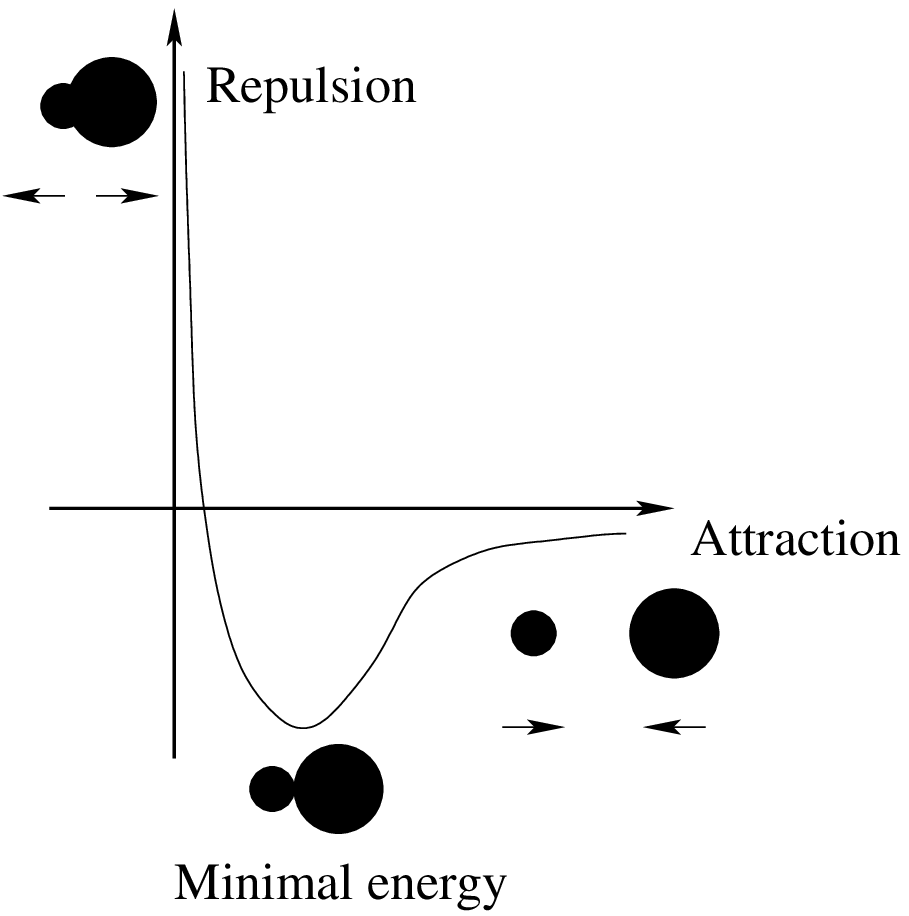}}
  \caption{Van der Waals interaction between two atoms.}
  \label{fig:VDW}
\end{figure}

There are two types of operators in a standard GA: crossover and mutation. The crossover mixes the information of two 
individuals, the parents, to create new individuals, the children. In our case, it 
swaps the ligand of two complexes. If the parent complexes are $S_1L_1$ and $S_2L_2$, the children complexes will be $S_1L_2$ and 
$S_2L_1$. It must be noticed that this type of operator can generate invalid complexes with atomic collisions. 
However, these complexes are penalized by the evaluation of the first objective. Its can be explained by one of the term of 
our first objective function: the Van der Waals term. Figure~\ref{fig:VDW} details the variation of the energy between two atoms
according to the distance of their center. 

There is an optimal distance that minimizes the energy, but if two atoms are too close the corresponding energy become very high. 
That is why our first objective function will penalize such ligand/site configuration. Thus, we do not need a mechanism to repair or 
check the generated complexes. 

Crossover operators do not add new information to the population. The parent information is just reordered in the children. 
Mutation adds some diversity in the new individuals just after applying the crossover. This unary operator is applied on an individual 
according to a global probability of mutation. Three mutation operators have been designed: rotation, translation and torsion rotation 
mutation. The rotation and translation operators provide rigid docking. The last operator adds some flexibility in the docking. If only 
one molecule can have its structure modified (typically the ligand) due to torsion rotation, it is a semi-flexible docking. If both
 molecules can be modified, it is a full flexible docking. In our case, we can make rigid, semi-flexible or full-flexible docking 
according to the configuration of our algorithm. The choice of mutation used depends on probabilities linked to each of them.

\subsubsection{Paralleling scheme}

In optimization methods, the evaluation step is resource consuming. Therefore, we use the well known master/slave 
paradigm for the individual evaluation. The master manages the GA and the slaves are used to evaluate one individual. In ParadisEO-PEO, 
the master is known as a \emph{runner} and another process called \emph{scheduler} dispatches the individuals that will be evaluated 
by the slaves. For instance, a parallel run with a master and ten slave needs in reality twelve processors.
 
\section{Results}

\subsection{Test protocol}

\subsubsection{test data}

In order to test our model, we use ligand/site complexes (LSC) from the CCDC/Astex data set. The original version of this data set is 
referenced in~\cite{BHV2007}. It corresponds to the benchmarking of the GOLD docking software. We have taken instances from the 
CCDC/Astex ``clean'' list. It corresponds to 224 diversified instances that suit well for docking benchmarking.

Table~\ref{table:instances} presents the first complexes taken from this list. 

\begin{table}[h]
  \centering
  \caption{Protein-ligand complexes used for benchmarking. PDB is the Protein Data Bank identifier of the complexes.}
  \label{table:instances}
  \begin{tabular}{|c|c|}
    \hline
    \textbf{Protein-ligand complexes} & \textbf{PDB}\\
    \hline
    Ribonuclease A / Uridine-2',3'-Vanadate & 6rsa\\
    HIV-1 Protease / G26 & 1mbi\\
    Thymidilate / CB3 & 2tsc\\
    HIV-1 Protease / G26 & 1htf\\
    Glucoamylase-471 / Alpha-d-mannose & 1dog\\
    \hline
  \end{tabular}
\end{table}

For the remaining of this article, the complexes will be designated by their corresponding Protein Data Bank identifier (PDB).
The docking algorithm is the last step of a larger work-flow of molecule/molecule interaction analysis: docking@GRID. According to 
this work-flow, we consider that the docking algorithm starts with two proteins, a ligand and another molecule with a potential 
binding site, in a stable conformation gained thanks to a folding algorithm. We also consider that the protein corresponding to 
the ligand is already in front of the binding site.

To prepare our instances, we have used the USCF Chimera software\footnote{www.cgl.ucsf.edu/chimera}. The
ligand has been manually extracted from its crystallographic location in order to have a \emph{seed} ligand. This seed is perturbed 
to generate a population of diversified individuals. These perturbations combine rotation, translation and torsion
 rotation. All these perturbations are applied randomly a given number of times (10 by default).

Table~\ref{table:seedDistance} details the deviation between the seed ligand used to initialize the GA population and the
ligand considered to be at the good location according to the crystallographic data. The computed deviation is the \textbf{R}oot 
\textbf{M}ean \textbf{S}quare \textbf{D}eviation (RMSD). According to~\cite{BT2008}, the RMSD is defined as followed:

\begin{equation}
  \label{RMSD}
  RMSD=\sqrt{\frac{\sum_{i=1}^{n}(dx_i^2+dy_i^2+dz_i^2)}{n}}
\end{equation}

In equation~\ref{RMSD}, $n$ corresponds to the total number of atoms. $dx_i$, $dy_i$ and $dz_i$ are the atomic coordinate
differences between the ligand predicted location and its location according to the crystallographic data. 

\begin{table}[h]
  \centering
  \caption{RMSD between the seed ligand and the ligand in its crystallographic location (according to the CCDC-Astex data set). 
The instances are cited according to their PDB identifier.}
  \label{table:seedDistance}
  \begin{tabular}{|c|c|}
    \hline
    \textbf{Instance} & \textbf{RMSD seed VS optimal (\AA)}\\
    \hline
    6rsa & 7.15\\
    1mbi & 7.93\\
    2tsc & 13.48\\
    1htf & 14.45\\
    1dog & 10.68\\
    \hline
  \end{tabular}
\end{table}

\subsubsection{Parameters}

Our population consists in 100 individuals. The probabilities of crossover and mutation are 0.9 and 0.5 respectively. In our
GA, the stopping criterion is a number of generations without improvement after a minimal number of generations. No improvement
means no new non-dominated solution discovery. In our tests, the minimal number of generations is 1000 and the number of generations 
without improvement is 100.

\subsubsection{Paralleling speed-up}

\begin{figure}[h]
    \centering{\includegraphics[scale=0.65]{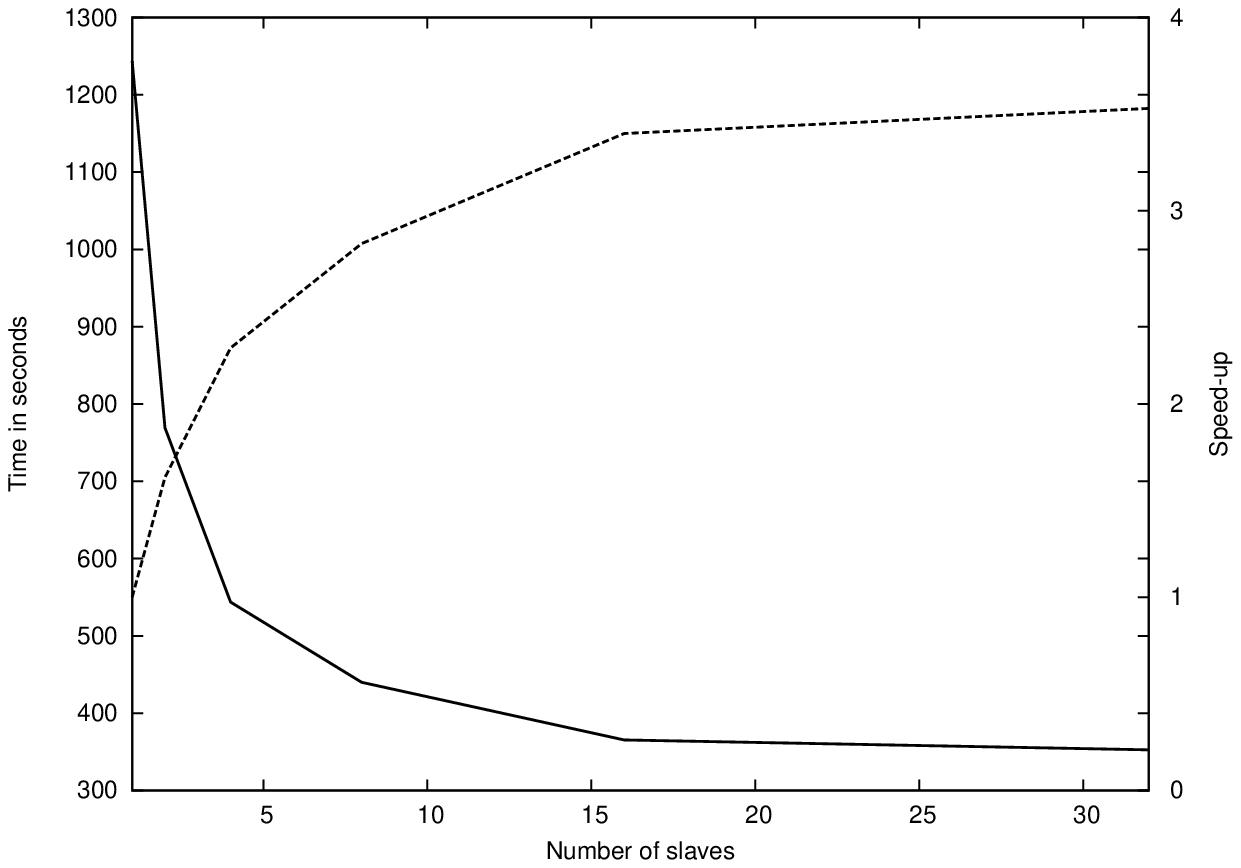}}
    \caption{Time (decreasing line) and speed-up (increasing line) according to the number of slaves used.}
    \label{fig:speedUp}
  \end{figure}

  Table~\ref{table:speed-up} and figure~\ref{fig:speedUp} shows an example of the speed-up obtained thanks the parallelization 
  of our GA for a small population of 32 individuals using Intel Xeon 3Ghz processors.  The speed-up corresponds to the ratio of the 
  time taken with one slave and the time with more slaves (2, 4, 8, 16 and 32 respectively). 
  
  \begin{table}[h]
    \centering
    \caption{Speed-up according to the number of slaves. Speed-up corresponds to time for 1 slave divided by the time of X slaves.}
    \label{table:speed-up}
    \begin{tabular}{|c|c|c|}
      \hline
      \textbf{Number of slaves} & \textbf{Time in seconds} & \textbf{Speed-up} \\
      \hline
      1 & 1243.76 & 1 \\
      2 & 769.30 & 1.62 \\
      4 & 543.73 & 2.29 \\
      8 & 439.85 & 2.83 \\
      16 & 365.32 & 3.4 \\
      32 & 352.61 & 3.53 \\
      \hline
    \end{tabular}
  \end{table}

  According to these data, we can establish that having a number of slaves equal to the population size is not necessarily an efficient 
  solution. It can be due to the time of communication between the master and a slave: packing an individual (master), send it 
  (from master to slave), unpacking the individual (slave), evaluate it (slave), packing the individual (slave), send it (from slave to 
  master) and unpacking the individual for using it (master). In an individual, the ligand coordinate vector is generally very small 
  ($<$ 100 atoms) but the binding site coordinate vector can be huge (more than 5000 atoms).
  
\subsection{Comparison}

All our tests have been run on a cluster of 64 Intel Xeon 3Ghz processors.
  
  \subsubsection{Performance Assessment}

  For each instance and each metaheuristic, a set of $10$ runs, with different initial populations, has been performed.
  In order to evaluate the quality of the non-dominated front approximations obtained for a specific test instance, we follow the 
  protocol given in~\cite{KTZ:06}. First, we compute a reference set~$Z_N^\star$ of non-dominated points extracted from the union of
  all these fronts. Second, we define $z^{max} = ( z^{max}_1 , z^{max}_2 )$, where $z^{max}_1$ (respectively $z^{max}_2$) denotes 
  the upper bound of the first (respectively second) objective in the whole non-dominated front approximations. Then, to measure 
  the quality of an output set~$A$ in comparison to $Z_N^\star$, we compute the difference between these two sets by using the unary 
  hypervolume metric~\cite{ZT:99}, $( 1.05 \times z^{max}_1 , 1.05 \times z^{max}_2 )$ being the reference point. The hypervolume 
  difference indicator computes the portion of the objective space that is dominated weakly by $Z_N^\star$ and not by~$A$.
  Furthermore, we also consider the unary additive $\epsilon$-indicator (I$^1_{\epsilon+}$) that gives the minimum value by which an
  approximation~$A$ has to be translated in the objective space to dominate weakly the reference set~$Z_N^\star$. As a consequence, 
  for each test instance, we obtain $10$~hypervolume differences and $10$~epsilon measures, corresponding to the $10$ runs, per algorithm.
  As suggested by Knowles et al.~\cite{KTZ:06}, once all these values are computed, we perform a statistical analysis on pairs of 
  optimization methods for a comparison on a specific test instance. To this end, we use the Mann-Whitney statistical test as described 
  in~\cite{KTZ:06}, with a p-value lower than $10\%$. Note that all the performance assessment procedures have been achieved using the
  performance assessment tool suite provided in PISA\footnote{The package is available at \url{http://www.tik.ee.ethz.ch/pisa/assessment.html}.}~\cite{BL+:03}.

  \begin{table}[t]
    \centering
    \caption{Comparison of different metaheuristics for the $Eps$ and the $I_H^-$ metrics by using a Mann-Whitney statistical test with a p-value of $5\%$. According to the metric under consideration, either the results of the algorithm located at a specific row are significantly better than those of the algorithm located at a specific column ($\succ$), either they are worse ($\prec$), or there is no significant difference between both ($\equiv$).} 
    \label{table:comparison}
    \begin{tabular}{|c|c|c|c|c|c|}
      \cline{3-6}
      \multicolumn{2}{c|}{ } & \multicolumn{2}{c|}{Eps} & \multicolumn{2}{c|}{$I_h^-$} \\
      \hline
      Instance & \emph{algorithms} & IBEA& NSGA II & IBEA & NSGA II \\
      \hline
      \multirow{2}{*}{6rsc} & IBEA & - &  $\succ$ &  - & $\succ$\\
      & NSGA II & $\prec$ &  - &  $\prec$ &  - \\
      \hline 
      \multirow{2}{*}{1mbi} & IBEA & - &  $\succ$ &  - &  $\succ$\\
      & NSGAII & $\prec$  &  - & $\prec$ &  -\\
      \hline 
      \multirow{2}{*}{2tsc} & IBEA &  -& $\equiv $ &  - &$\succ$ \\
      & NSGAII &$\equiv $ & - &$\prec$ &  - \\
      \hline
      \multirow{2}{*}{1htf} & IBEA  & - &  $\equiv $ &  - &  $\equiv $ \\
      & NSGAII 	& $\equiv $ & - &  $\equiv $ &   - \\
      \hline 
      \multirow{2}{*}{1dog} & IBEA & -  &  $\succ$ &  - &  $\succ$ \\
      & NSGA II & $\prec$ &  - & $\prec$ &  - \\
      \hline
    \end{tabular}
\end{table}

  According to table~\ref{table:comparison}, IBEA globally outperforms NSGA II for the instances of our problem.

\subsubsection{Docking results quality}

In order to evaluate our model, we have computed the RMSD of the ligand of our solutions with the crystallographic location
of the ligand. In the literature, it is common to estimate that a docking is good for a RMSD $\le 2.0$~\AA. Nevertheless,
the standard RMSD computation is not very robust according to several factor: size of the molecule, atoms used and not used, 
symmetric part, etc. So it is important to analyze well each solution for estimate its quality. Furthermore, the distance
between the initial solutions and the crystallographic solution is important because in most of the literature, this distance
is not $>$10~\AA ~and generally $\le 5$~\AA. 

Table~\ref{table:IBEA} summarizes the results of NSGA-II and IBEA on the five chosen instances. As the RMSD is not (and can not be) an 
objective of our model, all the archives generated during each run are analysed to know the quality of the encountered solutions. We have
remarked that the solutions with the best RMSD are not necessary in the final archive. It can be explained by a premature convergence of
our algorithm. In the same manner, one run makes on average of 225 000 evaluations. Comparing to other docking methods as 
Autodock (2 000 000 evaluations), it is not very high. Therefore, this number of evaluation can also significate a premature convergence
 of our algorithm. 

\begin{table}[h]
  \centering
  \caption{Best results for each instance with the NSGA-II and IBEA algorithms. For each algorithm the best RMSD and
the standard deviation (std) between the best RMSDs are given.}
  \label{table:IBEA}
  \begin{tabular}{|c|c|c|c|c|}
    \hline
    & \multicolumn{2}{|c|}{\textbf{NSGA-II best results}} & \multicolumn{2}{|c|}{\textbf{IBEA best results}}\\
    \hline
     \textbf{Instance} & RMSD (\AA) & std & RMSD (\AA) & std \\
    \hline
    6rsa & 1.66 & \textbf{1.04} & \textbf{1.32} & 1.3\\
    1mbi & 5.2 & \textbf{0.4} & \textbf{4.16} & 0.8\\
    2tsc & \textbf{2.19} & 2.75 & \textbf{2.19} & \textbf{2.68}\\
    1htf & 2.88 & 2.64 & \textbf{2.59} & \textbf{1.33}\\
    1dog & 4.38 & 0.99 & \textbf{2.44} & \textbf{0.56}\\
    \hline
  \end{tabular}
\end{table}

\begin{figure}[h]
  \centering{\includegraphics[scale=0.2]{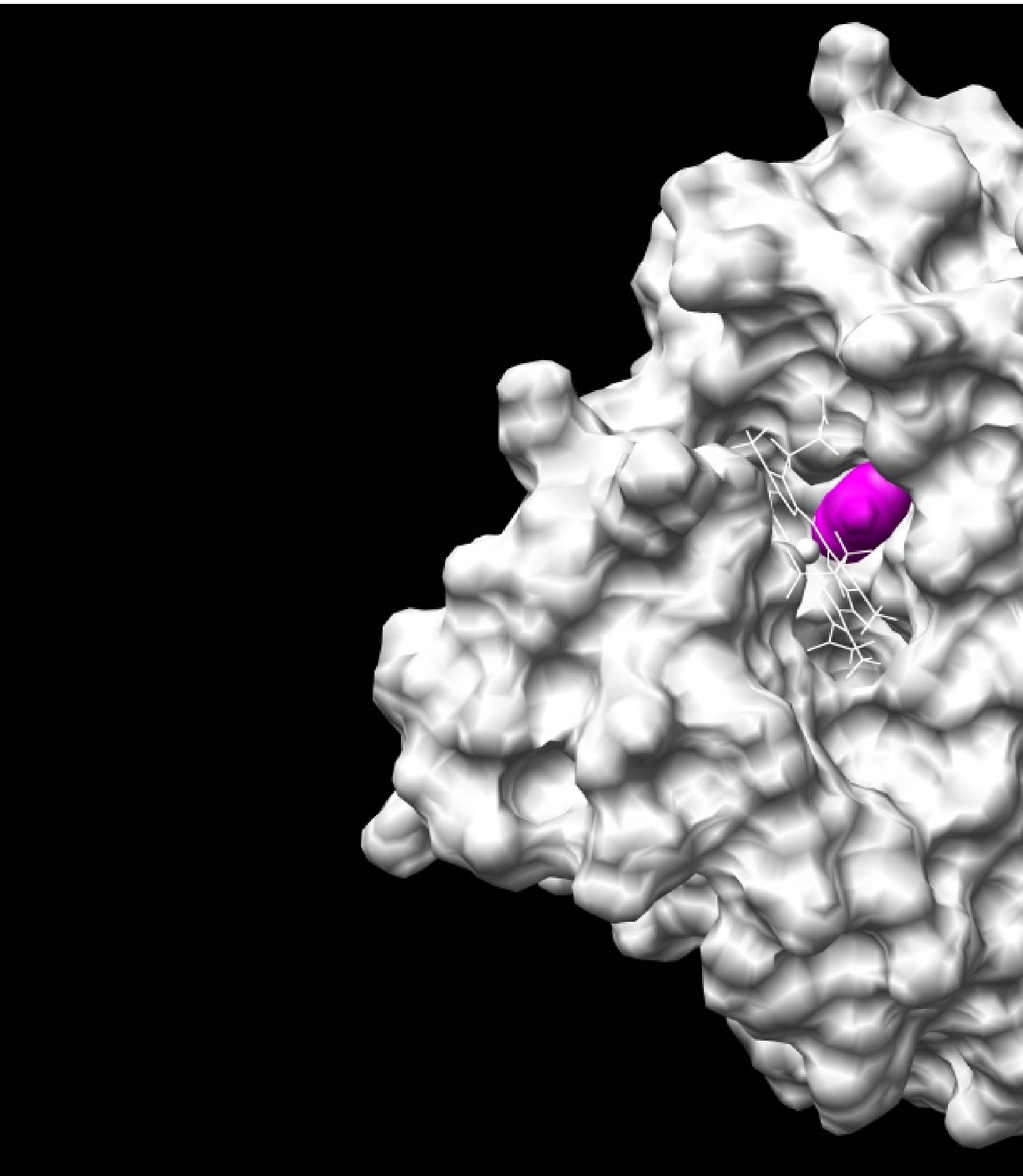}}
  \caption{Ligand/site complex coming from crystallographic data for the 1mbi instance.}
  \label{fig:1mbiOpt}
\end{figure}

According to the RMSD of our solution and the corresponding seed RMSD, we can estimate that our results are good for four instances 
(more particularly 6rsa, 2tsc and 1htf). Only 1mbi is problematic because the algorithm makes only few improvement of the RMSD (according
to the RMSD of the seed). An analysis of the 1mbi instance shows that the ligand is a very tiny molecule (9 atoms) that has to be put
in a big binding site (see Figure~\ref{fig:1mbiOpt}). Therefore, there are a lot of potential binding mode for the ligand, maybe of 
equivalent quality. 

According to the algorithm comparison, IBEA gives better or equivalent results on each instances. We can notice that the standard 
deviations are better for NSGA-II on 6rsa and 1mbi instances. This can be explained by the size of the instances because 6rsa and 1mbi are
the smallest instances of our dataset. 

IBEA has been already proved better than NSGA-II for several problems. Our results confirm this remark.

\begin{figure}[h]
  \centering{\includegraphics[scale=0.2]{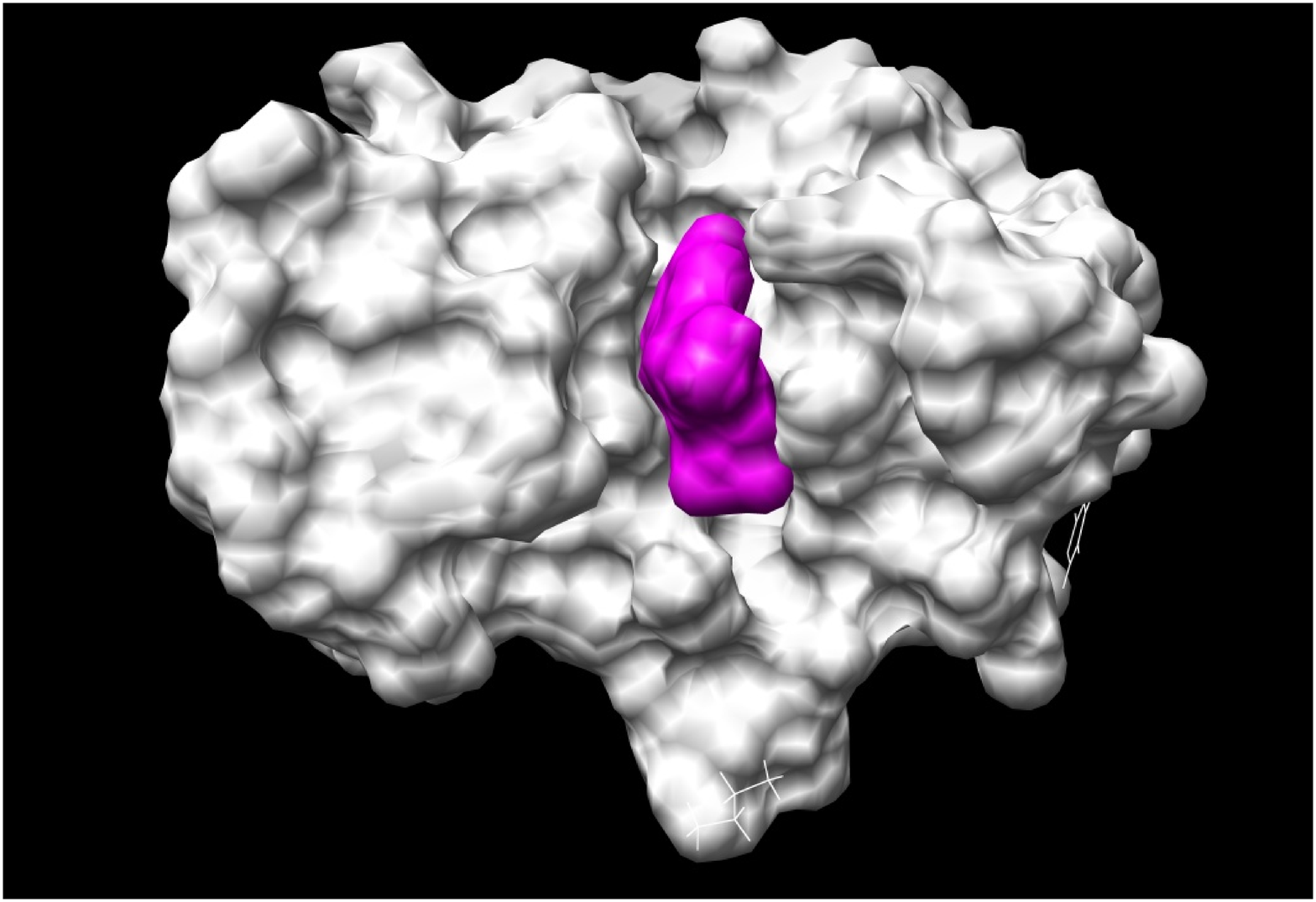}}
  \caption{Ligand/site complex coming from crystallographic data for the 6rsa instance.}
  \label{fig:6rsaOpt}
\end{figure}

\begin{figure}[h]
  \centering{\includegraphics[scale=0.2]{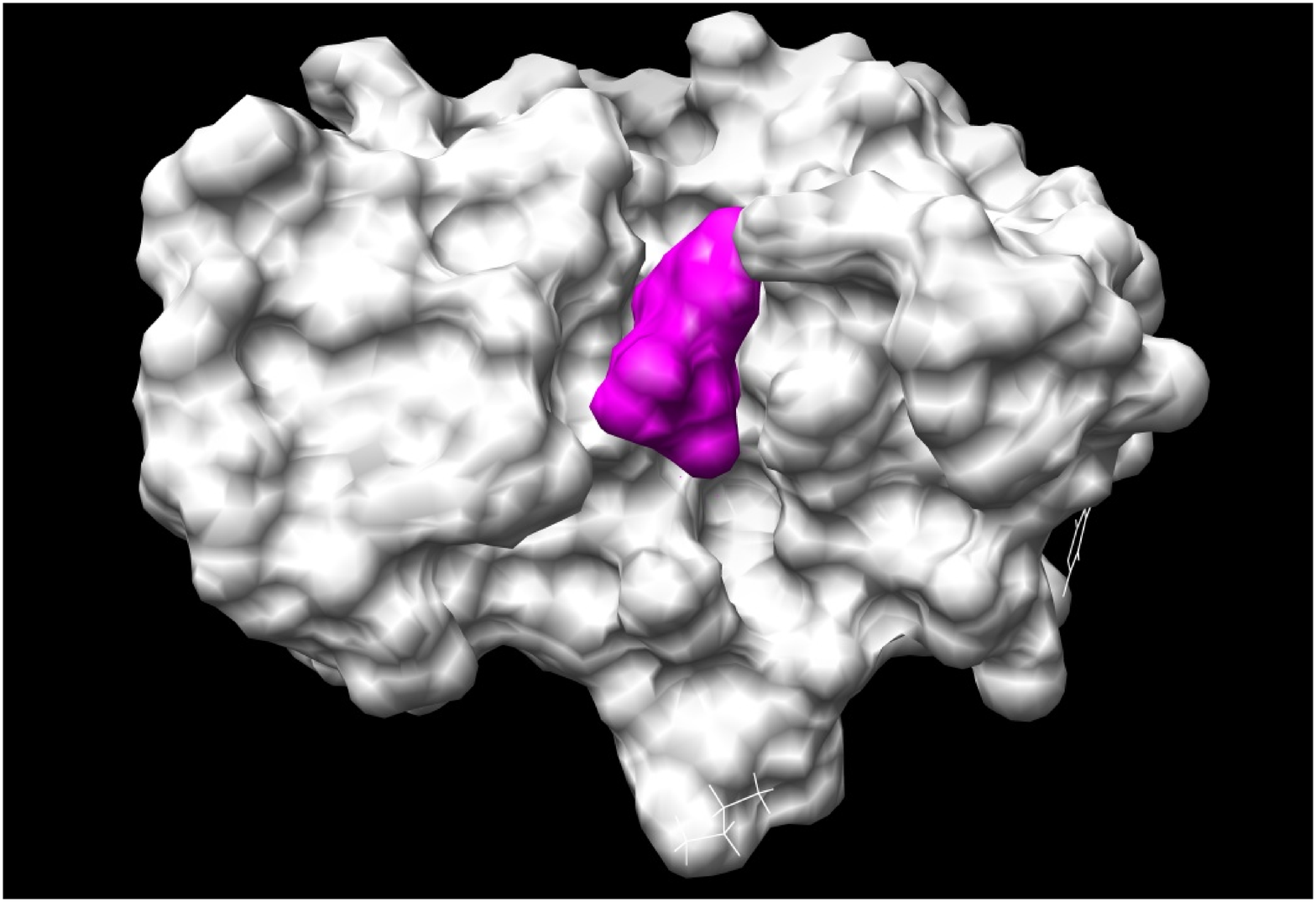}}
  \caption{Ligand/site complex coming from the individual having the best RMSD (1.66) with the NSGA-II algorithm.}
  \label{fig:6rsaBestNSGAII}
\end{figure}

\begin{figure}[h]
  \centering{\includegraphics[scale=0.2]{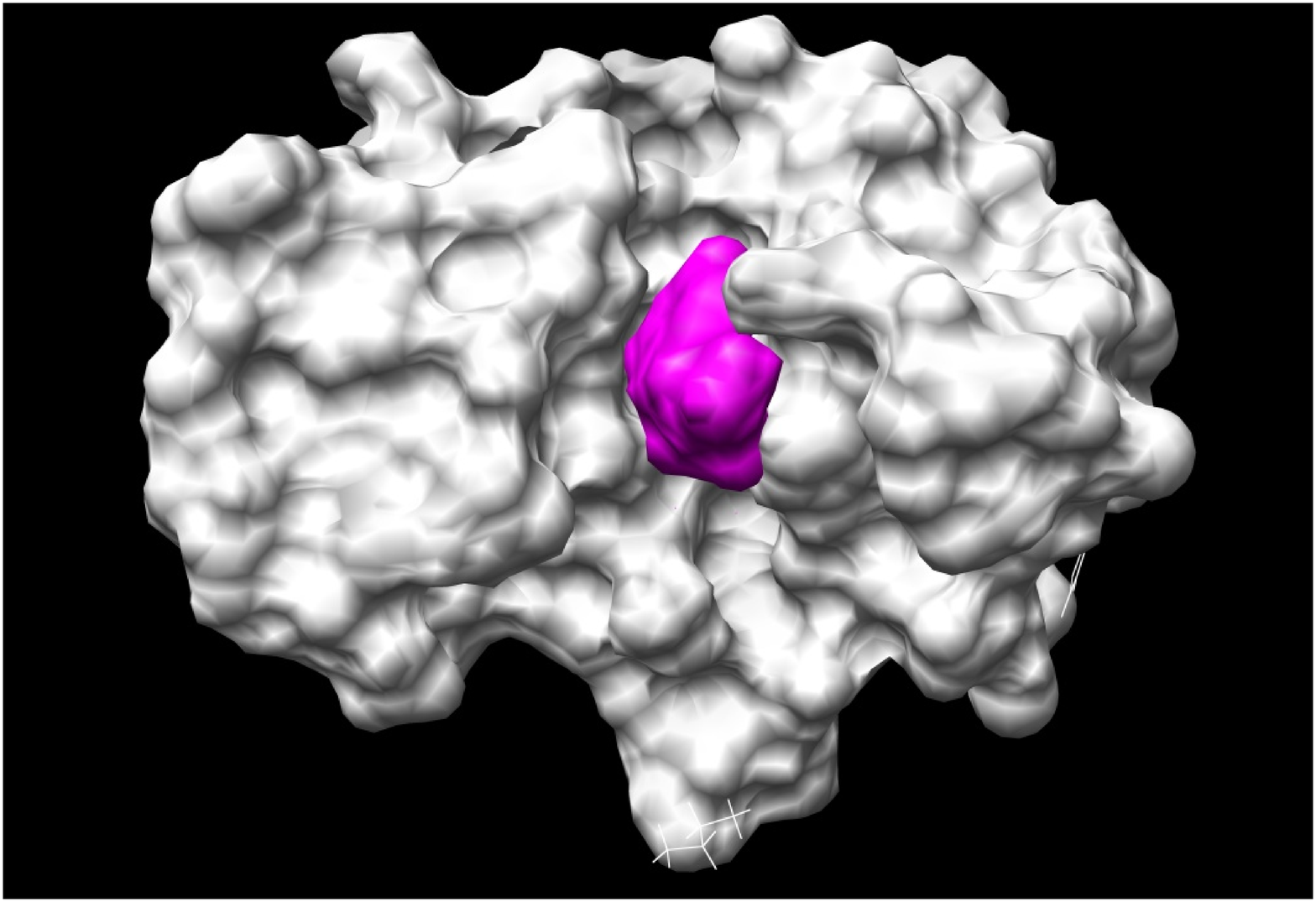}}
  \caption{Ligand/site complex coming from the individual having the best RMSD (1.32) with the IBEA algorithm.}
  \label{fig:6rsaBestIBEA}
\end{figure}

In order to compare visually a result of docking, the figure~\ref{fig:6rsaOpt} shows the crystallographic complex of the 6rsa instance.
Figure~\ref{fig:6rsaBestNSGAII} and figure~\ref{fig:6rsaBestIBEA} represent the complex with the minimal RMSD gained with the NSGA-II and
IBEA algorithms respectively.

NSGA-II proposes a ligand that is partially centered in the binding site. The ligand has not find its right conformation in the 
binding site. 

The IBEA solution has a lower RMSD because the ligand is better centered into the binding site.

Complementary tests are currently made in order to extend the number of instances tested and compare our approach with other
works of the literature. Nevertheless, according to our tests, our model has been validated and gives promising results.

\section{Conclusions}

In this article, a new bi-objective model for the molecular docking problem has been proposed. This model has been validated
thanks to instances of high confidence dedicated to docking benchmarking. Our model can be easily used with other energy function 
(and force field) and/or other molecular surfaces. A tri-objective version of our model is being tested. The
third objective is a robustness objective. It describes the quality of the ligand/site complex by making a sampling of
the energetic landscape around a current individual. However, this model is very time and resource consuming and has to
be improved in order to be used efficiently (grid computing). Furthermore, in order to improve the diversity of our population of 
solutions to prevent a potential premature convergence, new operators are planned to be added as the reverse mutation. The 
reverse mutation consists in making a big rotation of 180° of the ligand in order to increase the speed of convergence. This type of 
mutation can be useful in the case of a ligand well entered into the binding site but in the bad side so the associated RMSD is not low. 
In this case, it cannot be reversed by small rotations due to the lack of space. With the improvement of the algorithm behaviour,
we are getting a powerful docking method that will be available on-line throw the Docking@GRID platform (http://docking.futurs.inria.fr). 

\bibliographystyle{IEEEtran.bst}
\bibliography{bibliography}

\end{document}